\documentclass[aps,prl,showpacs,groupedaddress,twocolumn,preprintnumbers,amsmath,amssymb]{revtex4}

\usepackage{graphicx}
\usepackage{dcolumn}
\usepackage{bm}
\begin{document}

\title{Observation of Dipole-Induced Spin Texture in an $^{87}$Rb Bose-Einstein Condensate}

\author{Yujiro Eto$^{1}$}
\author{Hiroki Saito$^{2}$}
\author{Takuya Hirano$^{1}$}
\affiliation{%
$^{1}$Department of Physics, Gakushuin University, Toshima, Tokyo 171-8588, Japan\\
$^{2}$Department of Engineering Science, University of Electro-Communications, Chofu, Tokyo 182-8585, Japan}

\date{\today}
             
\begin{abstract}
We report the spin texture formation resulting from the magnetic dipole-dipole interaction in a spin-2 $^{87}$Rb Bose-Einstein condensate.
The spinor condensate is prepared in the transversely polarized spin state and the time evolution is observed under a magnetic field of 90 mG with a gradient of 3 mG/cm using Stern-Gerlach imaging.
The experimental results are compared with numerical simulations of the Gross-Pitaevskii equation, 
which reveals that the observed spatial modulation of the longitudinal magnetization is due to the spin precession in an effective magnetic field produced by the dipole-dipole interaction. 
These results show that the dipole-dipole interaction has considerable effects even on spinor condensates of alkali metal atoms.
\end{abstract}

\pacs{05.30.Jp, 03.75.Kk, 03.75.Mn, 67.85.Hj}
\maketitle
Dipolar interactions have attracted much attention in a wide variety of materials due to their long-range and anisotropic properties.
An example is the pattern formation in magnetic systems, in which the dipolar interactions, together with other interactions and geometries, form various patterns, such as stripes, bubbles, and vortices \cite{Seul95, Hurbert98, DeBell00}.
Microscopic control of dipolar particles can be applied to quantum information processing \cite{Jacksch00, DeMille02} and quantum simulations \cite{Micheli06}.

Recent experimental creation of Bose-Einstein condensates (BECs) of $^{52}$Cr \cite{Griesmaier05,Stuhler05,Lahaye08Nature,Lahaye08PRL,Beaufils08}, $^{164}$Dy \cite{Lu11}, and $^{168}$Er \cite{Aikawa12} atoms having 6-, 7-, and 10-$\mu_{B}$ magnetic dipole-moments, respectively ($\mu_{B}$ is the Bohr magneton), have stimulated theoretical and experimental studies of magnetic dipolar BECs \cite{Baranov08, Lahaya09}.
It is theoretically predicted that the interplay between the dipole interactions and spin degrees of freedom yields various intriguing phenomena, such as the Einstein-de Haas effect \cite{Kawaguchi06PRL1, Santos06, Gawryluk07, Gawryluk11} and ground state spin textures \cite{Yi06, Kawaguchi06PRL2}.
Experimentally the magnetization dynamics induced by the magnetic dipole-dipole interaction (MDDI) in spin-3 $^{52}$Cr BECs was observed, in which the external magnetic field is suppressed to below 1 mG so that the dipolar effects are not destroyed by Zeeman effects \cite{Pasquiou11}.
Although most spinor dipolar effects are typically obscured by Zeeman effects for a magnetic field of $\gtrsim$ 1 mG,
the weak MDDI in $^{87}$Rb, which has a magnetic moment of $\mu_B / 2$ or $\mu_B$, is expected to induce spin textures for specific spin preparation even in a magnetic field of about 100 mG \cite{Kawaguchi07}.
These spin textures originate from the spatially inhomogeneous spin precession in an effective magnetic field produced by the MDDI.

Spin texture formations in spinor BECs have been observed by several groups.
Spin domain structures have been developed by coherent spin exchange dynamics \cite{Kuwamoto04, Chang05} and quantum phase transitions through quenching of the quadratic Zeeman energy \cite{Sadler06, Bookjans11}.
The spontaneous formation of periodic spin patterns was observed in Ref. \cite{Kronjager10}.
The spontaneous decay of a helical spin structure to a modulated structure in Refs. \cite{Vengalattore08, Vengalattore10} may be ascribed to the MDDI, which is yet to be explained theoretically \cite{Kawaguchi10}.

\begin{figure}[tbp]
\includegraphics[width=8cm]{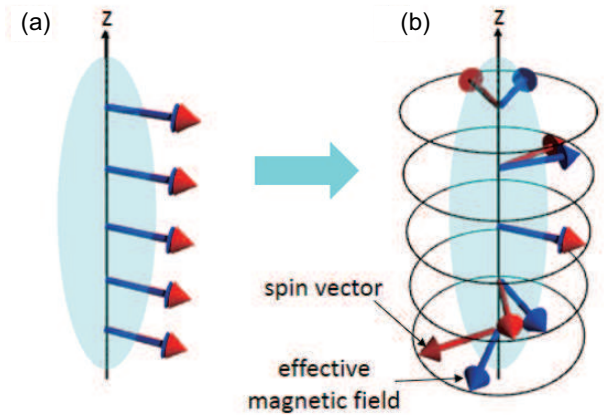}
\caption{
(color online) Schematic illustration of the distributions of the spin
vectors $\bm{f}$ and the effective magnetic field $\bm{b}_{\rm eff}$
produced by the magnetic dipoles.
The ellipsoids on the $z$ axis indicate the shape of the BEC.
(a) Initially, $\bm{f}$ and $\bm{b}_{\rm eff}$ have the same direction.
(b) By an external magnetic field gradient in the $z$ direction, $\bm{f}$
is twisted and $\bm{b}_{\rm eff}$ deviates from $\bm{f}$, which causes
precession of $\bm{f}$ around $\bm{b}_{\rm eff}$.
}
\label{f:schematic}
\end{figure}

In this Letter, we report the observation of spinor dipolar effects predicted in Ref. \cite{Kawaguchi07} using a spin-2 $^{87}$Rb BEC.
The magnetic moment of the spin-2 hyperfine state is twice as large as that of the spin-1 state.
In our scheme, the helical spin structure is created by Larmor precession subject to an external magnetic field of about 90 mG with an external field gradient of 3 mG/cm.
The helical spin state is then modulated by its own MDDI.
The time evolution of the spin distributions is observed using Stern-Gerach (SG) absorption imaging, and is then compared with the numerical simulation of the Gross-Pitaevskii (GP) equation with an MDDI.
The observed spatial modulation of the longitudinal magnetization is thereby identified as the effect of the spin precession in the effective magnetic field produced by the MDDI. 

The energy of the dipole-dipole interaction between magnetic dipoles $\bm{\mu}$ and $\bm{\mu}'$ located at $\bm{r}$ and $\bm{r}'$ has the form 
\begin{equation} 
\label{Eddi}
\frac{\mu_0}{4\pi |\bm{r} - \bm{r}'|^3} \left[ \bm{\mu} \cdot \bm{\mu}'
- 3 (\bm{\mu} \cdot \bm{e}) (\bm{\mu}' \cdot \bm{e}) \right],
\end{equation}
where $\mu_0$ is the magnetic permeability of the vacuum and $\bm{e} = (\bm{r} - \bm{r}') / |\bm{r} - \bm{r}'|$.
When an external magnetic field $B_z$ is applied in the $z$ direction and the Larmor precession with frequency $\mu B_z / \hbar$ is much faster than the other characteristic dynamics, we can take a time average of Eq.~(\ref{Eddi}), giving~\cite{Giovanazzi02,Kawaguchi07}
\begin{equation} 
\label{Eavg}
\frac{\mu_0 (1 - 3 e_z^2)}{8\pi |\bm{r} - \bm{r}'|^3}
\left( 3 \mu_z \mu_z' - \bm{\mu} \cdot \bm{\mu}' \right),
\end{equation}
which is the effective MDDI observed in this Letter.

In the mean-field theory for BECs, the magnetic dipole density is described by $g \mu_{\rm B} \bm{f} = g \mu_{\rm B} \sum_{m_F, m_F'} \psi_{m_F}^* \bm{S}_{m_F, m_F'} \psi_{m_F'}$, where $g$ is the Land\'e $g$ factor for the hyperfine spin, 
$\psi_{m_F}$ is the macroscopic wave
function ($m_{F} = -2, -1, \cdots, 2$), and $\bm{S}$ is the vector of spin-2 matrices.
From Eq.~(\ref{Eavg}), the mean-field energy of the Larmor-averaged dipoles is written as $E_{\rm ddi} = -\int g \mu_{\rm B} \bm{f} \cdot
\bm{b}_{\rm eff} d\bm{r}$, where
\begin{equation} 
\label{beff}
\bm{b}_{\rm eff} = \frac{\mu_0 g \mu_{\rm B}}{8\pi} \int d\bm{r}'
\frac{1 - 3 e_z^2}{|\bm{r} - \bm{r}'|^3} \left[ 3 f_z(\bm{r}') \hat z
- \bm{f}(\bm{r}') \right]
\end{equation}
is the effective magnetic field produced by the dipoles,
with $\hat z$ being the unit vector in the $z$ direction.

Let us consider a situation in which all spin vectors are aligned in the $x$ direction. 
It follows from Eq.~(\ref{beff}) that the effective magnetic field $\bm{b}_{\rm eff}$ has the same direction as the spin vectors $\bm{f}$~\cite{Kawaguchi07}, as illustrated in Fig.~\ref{f:schematic}(a), 
and hence the Larmor precession around $\bm{b}_{\rm eff}$ does not change the spin direction.
Applying a magnetic field gradient $dB_z / dz$, we can twist the spin vectors along the $z$ axis~\cite{Higbie05}.
In such a helical spin structure, $\bm{b}_{\rm eff}$ deviates from $\bm{f}$, as depicted in Fig.~\ref{f:schematic}(b).
As a result, $\bm{b}_{\rm eff}$ causes spin precession that depends on the position $\bm{r}$, and a spin pattern is expected to be formed.

\begin{figure}[tbp]
\includegraphics[width=8cm]{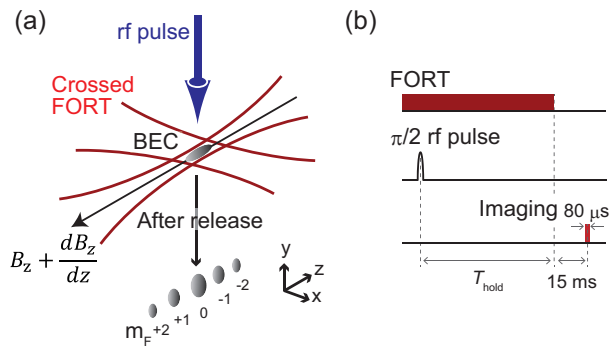}
\caption{
(color online) (a) Schematic illustration of the experimental setup.
The BEC is confined in the crossed FORT and an rf pulse prepares the initial spin state as shown in Fig. 1(a).
After a hold time $T_{\mathrm{hold}}$, the atoms are released from the FORT and the spin components are separated by the SG method.
(b) Timing diagram for texture formation and its measurement. 
The envelope of $\pi$/2 rf pulse has a Gaussian shape with a standard deviation of 58 $\mu$s.
}
\label{f:experiment}
\end{figure}

The outline of the experimental setup and the timing diagram are shown in Figs. \ref{f:experiment}(a) and \ref{f:experiment}(b), respectively.
We produce an $^{87}$Rb BEC containing $3.9(2) \times10^5$ atoms in the hyperfine state $| F = 2, m_{F} = 2 \rangle$ in a crossed far-off-resonant optical dipole trap (FORT) with axial and radial frequencies of $\omega_z / (2\pi) = 20$ Hz and $\omega_r / (2\pi) =120$ Hz (see Ref. \cite{Eto13PRA} for a more detailed description).
In order to control the external magnetic field, the whole experimental setup is installed inside a magnetic shield room whose walls consist of permalloy plates.
The external magnetic field of $B_z = 92.6$ mG with a gradient of $dB_z / dz = 3$ mG/cm is aligned with the axis of the trap ($z$ direction).

The transversely polarized spin state is prepared by applying a resonant $\pi/2$ radio-frequency (rf) pulse, 
and thereby the $z$-dependent Larmor precession in the $x$-$y$ plane induces spin dynamics.
After holding for a variable time, $T_{\mathrm{hold}}$, the BECs are released from the FORT.
Each $m_F$ component is spatially separated along the $z$ direction by the SG method. 
After a time-of-flight (TOF) of 15 ms, the atomic distribution of each $m_{F}$ component is measured using absorption imaging.

\begin{figure}[tbp]
\includegraphics[width=8cm]{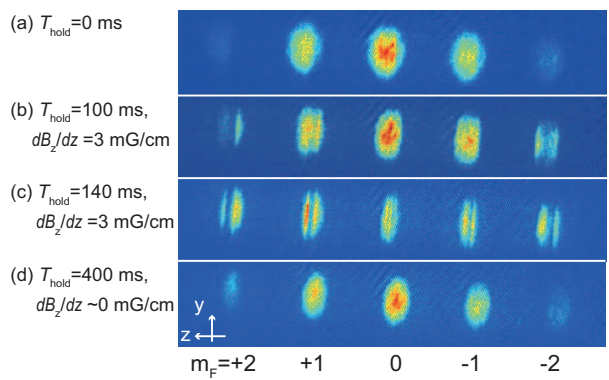}
\caption{
(color online) Absorption images of condensates taken at (a) $T_{\mathrm{hold}} = 0$ ms, (b) $T_{\mathrm{hold}} =100$ ms, (c) $T_{\mathrm{hold}} =140$ ms, and (d) $T_{\mathrm{hold}} =400$ ms.
In (a), (b) and (c), the magnetic field gradient of $dB_z / dz = 3$ mG/cm is applied along the $z$-direction.
In (d), the magnetic field gradient is almost zero.
}
\label{f:absorption}
\end{figure}

The texture formation is clearly observed, as shown in Figs. \ref{f:absorption}(a)-\ref{f:absorption}(c).
The double peaks are generated along the $z$ direction in $m_F \neq 0$ components as $T_{\mathrm{hold}}$ is increased.
In contrast, when $dB_{z}/dz$ is almost zero, 
in which $\bm{b}_{\rm eff}$ always has the same direction as the spin and does not affect the spin dynamics [see Fig. \ref{f:schematic}(a)], 
apart from the decrease in atomic number due to the photon scattering from the trap light, no clear changes are observed, even with $T_{\mathrm{hold}} = 400$ ms [Fig. \ref{f:experiment}(d)].
For a TOF of 15 ms, the atomic distributions in the absorption images, such as the double peaks in Fig. 3, reflect the spatial distributions in the FORT rather than the momentum distributions \cite{Eto13APEX}.

\begin{figure}[t]
\includegraphics[width=8cm]{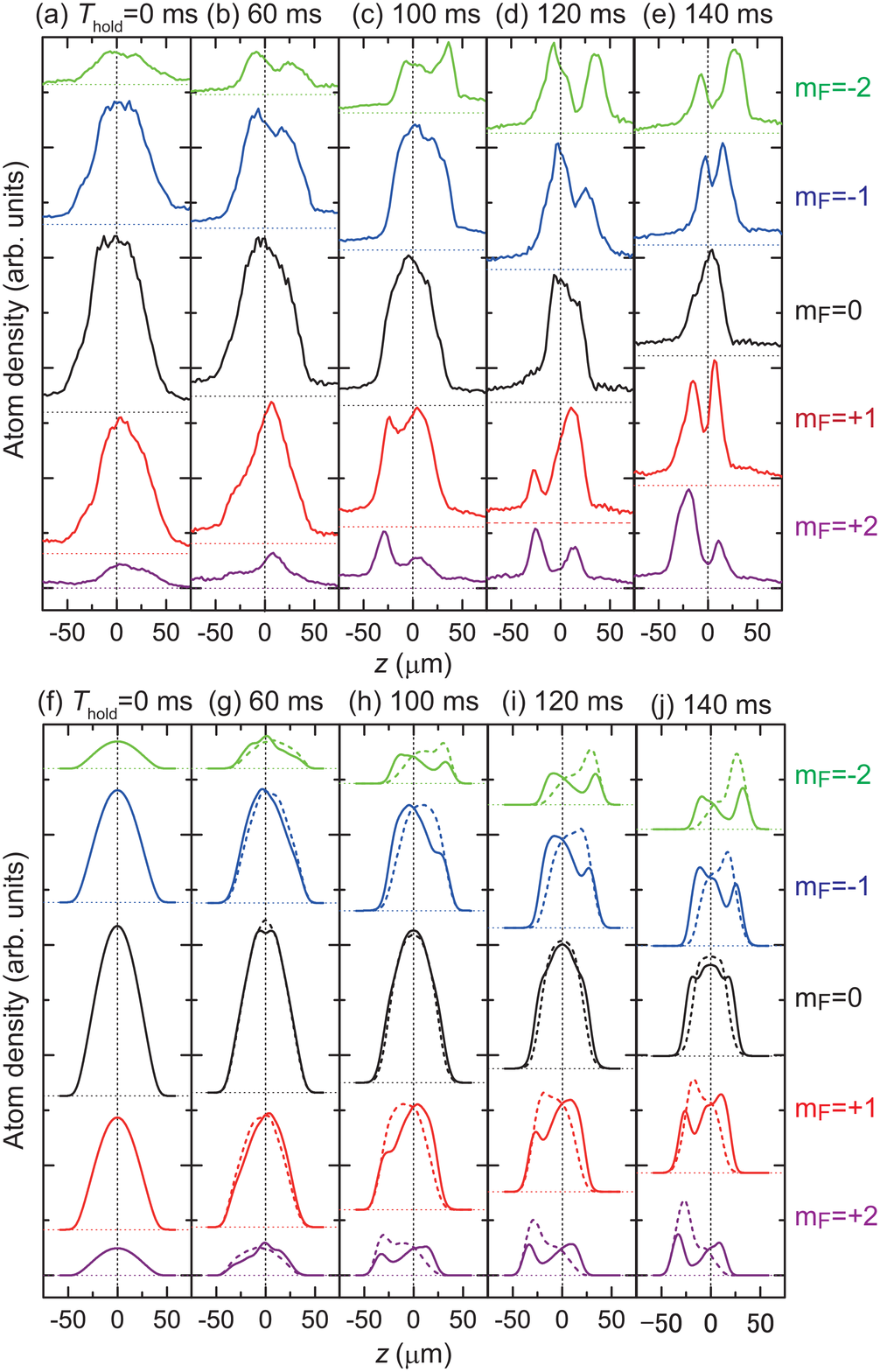}
\caption{(color online)  
(a)-(e) Experimentally observed atomic distributions. The absorption images are integrated over $y$. 
The distances traveled by each component during the SG measurement are subtracted from $z$.
(f)-(j) Numerically obtained atomic distributions. The density is integrated over $x$ and $y$. 
The solid and dashed curves indicate the results with and without the MDDI.}
\label{f:od}
\end{figure}


In order to investigate the effect of the MDDI on the spin texture formation, we numerically solve the three dimensional GP equation,
\begin{equation} \label{GP}
	i\hbar \frac{\partial \psi_{m_{F}}(\bm{r}, t)}{\partial t} =
\frac{\delta E}{\delta \psi_{m_{F}}^*(\bm{r}, t)},
\end{equation}
where the right-hand side stands for the functional derivative.
The mean-field energy $E$ in Eq.~(\ref{GP}) has the form,
\begin{equation} \label{E}
E = \int d\bm{r} \sum_{m_{F}} \psi_{m_{F}}^* \left[ -\frac{\hbar^2}{2M} \nabla^2 +
V_{m_{F}}(\bm{r}) \right] \psi_{m_{F}} + E_{\rm s} + E_{\rm ddi},
\end{equation}
where $M$ is the mass of $^{87}{\rm Rb}$ and
$V_{m_{F}} = M [\omega_r^2 (x^2 +	y^2) + \omega_z^2 z^2] / 2 + m_{F} \mu_{\rm B} B_z'(z) / 2 -(m_F \mu_B B_z)^2 / (4E_{\rm hf})$ with $E_{\rm hf}$ being the hyperfine splitting energy. 
The uniform linear Zeeman term is eliminated from  Eq.~(\ref{E}) in the spin space rotating at the Larmor frequency.
In the TOF stage, the harmonic potential in $V_{m_{F}}$ is switched off.
The $s$-wave interaction energy in Eq.~(\ref{E}) is given by 
$E_{\rm s} = \int d\bm{r} 4 \pi \hbar^2 (b_0 \rho^2 + b_1 \bm{f}^2 + b_2 |A_0|^2) / (2M)$, 
where
$b_0 = (4 a_2 + 3 a_4) / 7$, $b_1 = (a_4 - a_2) / 7$, $b_2 = (7 a_0 - 10 a_2 + 3 a_4) / 7$ 
with $a_f$ being the $s$-wave scattering length with the colliding channel of total spin $f$, $\rho = \sum_{m_{F}} |\psi_{m_{F}}|^2$, and $A_0
= (2\psi_2 \psi_{-2} - 2 \psi_1 \psi_{-1} + \psi_0^2) / \sqrt{5}$.
The initial state is prepared by the imaginary-time propagation method, and the time evolution is obtained by the pseudo-spectral method, where
the convolution integral in the MDDI is calculated using a fast Fourier transform.
The atomic loss due to the inelastic collision of $F = 2$ atoms hardly affects the dynamics and is neglected.


\begin{figure}[tbp]
\includegraphics[width=8cm]{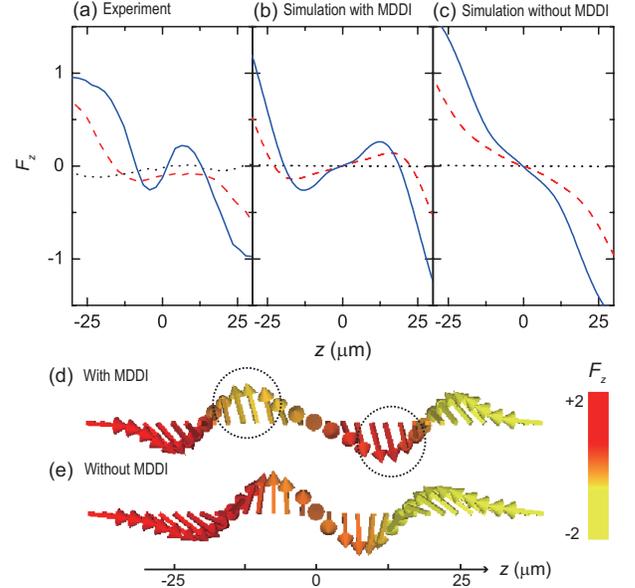}
\caption{(color online) The spatial distributions of spin orientation.
(a)-(c) $F_{z}(z)$ at  $T_{\mathrm{hold}} = 0$ ms (dotted curves), $100$ ms (dashed curves), and $140$ ms (solid curves). 
(a) is calculated from the experimental results in Figs. \ref{f:od}(a), \ref{f:od}(c), and \ref{f:od}(e).
(b) and (c) are the numerical results with and without the MDDI, respectively.
(d) and (e) are the numerically obtained spin vectors on the $z$ axis at  $T_{\mathrm{hold}} = 140$ ms.
The color represents the magnitude of $F_{z}$.
The dotted circles mark where the effect of the MDDI is significant.
}
\label{f:magnetization}
\end{figure}

The experimentally observed and numerically simulated atomic distributions of each $m_{F}$ component at various $T_{\mathrm{hold}}$ are shown in Figs. \ref{f:od}(a)-\ref{f:od}(e) and \ref{f:od}(f)-\ref{f:od}(j), respectively.
The distances traveled by each component during the SG measurement are subtracted from $z$.
The features of the experimental results including the double peak structures are well reproduced by the numerical results with the MDDI [the solid curves in Figs. \ref{f:od}(f)-(j)].
On the other hand, when the MDDI is not included in the GP equation [dashed curves in Figs. \ref{f:od}(f)-\ref{f:od}(j)],
the experimentally observed double peak structures cannot be reproduced.
The left (right) side peaks in the $m_{F} = +1$ and $+2$ ($-1$ and $-2$) components are due to the spin current generated by the magnetic field gradient,
and the other peaks originate from the MDDI.


The longitudinal magnetization is obtained from the data in Fig. \ref{f:od} by using
\begin{equation}
F_{z}(z) = \frac{\sum_{m_{F}} m_{F} N_{m_{F}}(z)}{\sum_{m_{F}} N_{m_{F}}(z)},
\label{eq1}
\end{equation}
where $N_{m_{F}}(z)$ is the atom number density in component $m_{F}$ integrated over $x$ and $y$.
Figures \ref{f:magnetization}(a)- \ref{f:magnetization}(c) show the $z$ dependence of $F_{z}(z)$ for $T_{\mathrm{hold}} = 0$ ms (dotted curves), $100$ ms (dashed curves), and $140$ ms (solid curves). 
The spatial modulations of $F_{z}(z)$ are clearly observed in the experimental data  [Fig. \ref{f:magnetization} (a)] and the simulation with the MDDI [Fig. \ref{f:magnetization} (b)].
In the numerical simulation without the MDDI [Fig. \ref{f:magnetization}(c)], on the other hand, $F_{z}(z)$ is monotonically decreased with $z$ due to the spin current generated by the magnetic gradient force.

The double-peak structures in Fig. \ref{f:od} and the modulation of $F_z(z)$ in Fig. \ref{f:magnetization} can be understood from the spin dynamics. 
Figure \ref{f:magnetization}(d) shows the spin vector distribution $\bm{F}(\bm{r}) = \bm{f}(\bm{r}) / \rho(\bm{r})$ along the $z$ axis obtained by the numerical simulation with the MDDI. 
The spin orientation twisted by the magnetic field gradient produces the effective magnetic field $\bm{b}_{\rm eff}$ as shown in Fig. \ref{f:schematic}(b). 
As a result of the Larmor precession around $\bm{b}_{\rm eff}$, the spin vectors acquire the $+z$  ($-z$) component for $z > 0$ ($z < 0$), which is marked by the dotted circles in Fig. \ref{f:magnetization}(d). 
At the same time, the spin current generated by the magnetic field gradient accumulates the $\pm z$ spin components at the $\mp z$ edges of the BEC. 
Thus, there appear two regions of $F_z > 0$ (red, dark gray) and those of $F_z < 0$ (yellow, light gray) in Fig. \ref{f:magnetization}(d), which is the origin of the double-peak structures in Fig. \ref{f:od} and the modulation in Figs. \ref{f:magnetization}(a) and \ref{f:magnetization}(b).
From the simulation, the magnitude of $\bm{b}_{\rm eff}$ is found to be in the order of 10 $\mu$G.
Figure \ref{f:magnetization}(e) shows the spin vector distribution $\bm{F}(\bm{r})$ obtained by the simulation without the MDDI, in which $F_z$ monotonically decreases with $z$.

A possible reason of the quantitative differences between the experimental and theoretical results in Figs. \ref{f:od} and \ref{f:magnetization} is a finite temperature effect.
The thermal fraction of about 10\% causes diffusion of each spin component, which affects the spin texture. 
In fact, Higbie \textit{et al.} reported that the thermal diffusion drastically reduced the spin coherence under a magnetic field gradient \cite{Higbie05}. 
In addition, the trap potential has slight asymmetry in the axial direction due to the experimental imperfection.
This would also cause the deviations between the experimental and theoretical results.

In conclusion, we reported the observation of spinor dipolar effects in an $^{87}$Rb BEC, in which the effective magnetic field induced by the MDDI forms the modulated helical spin texture.
The observation is in good agreement with the numerical simulation of the GP equation including an MDDI.
These experimental results show that MDDIs have considerable effects on the BECs of $^{87}$Rb for specific spin states even though the isotropic contact interaction and Zeeman energies dominate the MDDI energy. 

We would like to thank M. Sadgrove for his valuable comments.
This work was supported by the Japan Society for the Promotion of Science (JSPS) through its Funding Program for World-Leading Innovation R\&D in Science and Technology (FIRST Program), 
a Grant-in-Aid for Scientific Research (C) (No. 23540464) and a Grant-in-Aid for Scientific Research on Innovation Areas Fluctuation \& Structure (No. 25103007) from the Ministry of Education, Culture, Sports, Science, and Technology of Japan.

\end{document}